\documentclass[aps,prl,reprint,superscriptaddress,showpacs]{revtex4-1}
\usepackage{graphicx}% Include figure files
\usepackage{amssymb}
\usepackage{amsmath}
\usepackage{bm}
\usepackage{verbatim}
\usepackage[bookmarks=false,colorlinks=true,urlcolor=blue,citecolor=blue,linkcolor=blue]{hyperref}
\usepackage[normalem]{ulem}

\usepackage[bbgreekl]{mathbbol}

\renewcommand{\bf}[1]{{\textbf #1}}

\begin{document}

\title{Nematic Bond Theory of Heisenberg Helimagnets}

\author{Michael Schecter}
\affiliation{Center for Quantum Devices, Niels Bohr Institute, University of Copenhagen, 2100 Copenhagen, Denmark}
\author{Olav F. Sylju{\aa}sen}
\affiliation{Department of Physics, University of Oslo, P. O. Box 1048 Blindern, N-0316 Oslo, Norway}
\author{Jens Paaske}
\affiliation{Center for Quantum Devices, Niels Bohr Institute, University of Copenhagen, 2100 Copenhagen, Denmark}

\date{\today}

\begin{abstract}
We study classical two-dimensional frustrated Heisenberg models with generically incommensurate groundstates. A new theory for the spin-nematic ``order by disorder" transition is developed based on the self-consistent determination of the effective exchange coupling bonds. In our approach, fluctuations of the constraint field imposing conservation of the local magnetic moment drive nematicity at low temperatures. The critical temperature is found to be highly sensitive to the peak helimagnetic wavevector, and vanishes continuously when approaching rotation symmetric Lifshitz points. Transitions between symmetry distinct nematic orders may occur by tuning the exchange parameters, leading to lines of bicritical points.
\end{abstract}


\maketitle

Spatial winding of spin magnetization $\--$ helimagnetism $\--$ is an intriguing magnetic effect frequently arising from strongly competing exchange couplings \cite{Yoshimori59, Villain59,Kaplan59} or long-range, oscillatory indirect exchange couplings \cite{Jensen91}. A generic helimagnetic groundstate 
%spin configuration $\textbf{S}_{\textbf{Q}}$ with wavevector $\textbf{Q}\neq \textbf{G}/2$, where $\textbf{G}$ %is a reciprocal lattice vector, 
breaks both the continuous spin-rotation and the discrete lattice rotation symmetries \cite{Villain77}. In two spatial dimensions, the Mermin-Wagner theorem precludes continuous symmetry breaking at any finite temperature $T$. This permits the remarkable possibility of a nematic spin transition at $T_c>0$, into a state with broken lattice-rotation symmetry, but restored spin-rotation symmetry \cite{Villain77,Henley89,Chandra90}. 

In the case of collinear antiferromagnets (CAF), previous studies \cite{Henley89,Chandra90,Fang08} have predicted nematic order based on entropic gains from linear spin-wave fluctuations around classically degenerate magnetically ordered states (see also \cite{Xu08,Fernandes12}). This ``order by disorder" scenario is specific to the CAF phase since it requires an internal $O(3)$ groundstate degeneracy (in addition to global $O(3)$ rotations) that only occurs for special commensurate values of the helimagnetic wavevector \cite{Villain77}. Despite strong numerical evidence for the existence of nematic order in \emph{incommensurate} two-dimensional helimagnets \cite{Weber03,Capriotti04,Seabra16}, its theoretical understanding and the conditions under which it occurs, remains elusive.

This Letter develops a new approach to the spin-nematic problem based on the self-consistent determination of the effective exchange coupling bonds (exchange bonds, for brevity). Within this nematic bond theory, a spontaneous nematic distortion of the exchange bonds $J_\bf{q}^{\rm eff}\sim \chi^{-1}_\bf{q}$, where $\chi_\bf{q}$ is the spin susceptibility in momentum space, become stabilized at low temperatures due to the ``order by disorder" mechanism. However, contrary to the conventional spin-wave mechanism \cite{Henley89,Chandra90}, in our theory the nematic transition is driven by fluctuations of the constraint field $\lambda_j$ that imposes the conservation of the local magnetic moment $|\bf{S}_j|^2=1$. This framework respects the Mermin-Wagner theorem, and describes systems with arbitrary commensurate/incommensurate spin correlations and arbitrary groundstate degeneracy. We are thus able to study nematic order near Lifshitz transitions in the spin susceptibility $\chi_\bf{q}$, where the peak wavevector cannot remain commensurate due to the sign change of the quadratic coefficient. 
We find that when the wavevector $\bf{Q}_0$ at the Lifshitz point is rotation invariant, e.g. $\textbf{Q}_0=(0,0)$ or $\bf{Q}_0=(\pi,\pi)$ for the square lattice, the critical temperature vanishes as $T_c\propto |\bf{Q}-\bf{Q}_0|^2$, as $|\bf{Q}-\bf{Q}_0|\to0$.

\begin{figure}[t]
\includegraphics[width=.7\columnwidth]{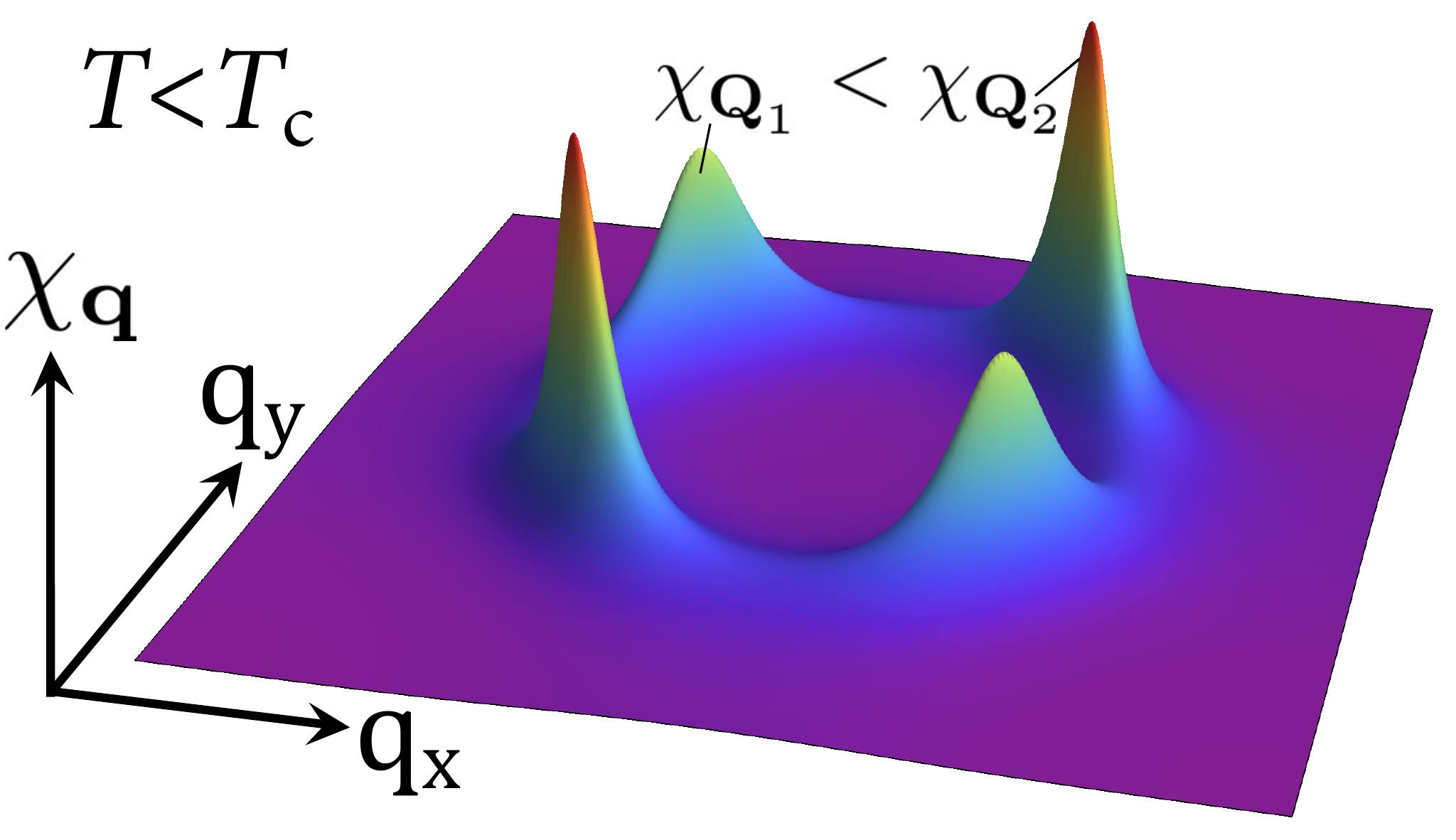}
\caption{(Color online) Schematic of the spin susceptibility $\chi_\bf{q}$ in the ordered phase with broken $C_{4v}$ symmetry. Here nematic order preserves mirror reflections about the diagonal ($q_x=q_y$) and anti-diagonal ($q_x=-q_y$).}
\label{fig:1}
\end{figure}

As a concrete example, we apply our theory to the square lattice ferromagnetic $J_1$-$J_2$-$J_3$ Heisenberg model. In this case, nematic order can break $C_{4v}$ symmetry in two distinct ways, depending on how the order parameter transforms under mirror reflections. In Fig.~\ref{fig:1} we show the case where the susceptibility has broken rotation symmetry but is symmmetric under mirror reflections about the diagonals. The second case of axis mirror symmetry occurs when the peaks lie along the $q_x,q_y$ axes, and can be achieved by tuning the exchange parameters. In this way one can cross through finite temperature nematic bicritical points, associated with the intersection of symmetry distinct nematic critical surfaces. 
%%%%%%%%%%%
%%%%%%%%%%%
%%%%%%%%%%%
%%%%%%%%%%%

We focus on systems described by the helimagnet Heisenberg Hamiltonian
\begin{equation}
\label{eq:H}
H=\sum_\mathbf{q}J_\mathbf{q}|\mathbf{S}_\mathbf{q}|^2.
\end{equation}
Here the spin variables $\mathbf{S}_{\textbf q}$ are written in reciprocal space and are assumed to be classical vectors with $N=3$ components and unit norm on every site $i$ of the direct lattice, $|\textbf{S}_i|^2=1$. The exchange coupling $J_{\textbf{q}}$ respects the symmetries of the lattice, and leads to a helimagnetic groundstate when the minima of $J_{\textbf{q}}$ form a set $\textbf{q}=\textbf{Q}_\alpha$ with more than one distinct element $\alpha$. In this case, the groundstate manifold (excluding $O(3)$ rotations) generically exhibits a \emph{discrete} degeneracy, associated with the states $\textbf{S}_{\alpha i}=\textbf{u}\cos \textbf{Q}_\alpha\cdot\textbf{R}_i+\textbf{v}\sin \textbf{Q}_\alpha\cdot\textbf{R}_i$  \cite{Villain77}, where $\textbf{u},\,\textbf{v}$ are orthonormal vectors and $\textbf{R}_i$ are lattice sites. We must stress that states formed by taking linear combinations of $\bf{S}_{\alpha i}$ are generically \emph{excluded} by the local constraint $|\textbf{S}_i|^2=1$ (an important exception is the CAF \cite{Villain77,Henley89,Chandra90}). Thus, the conventional spin-wave ``order by disorder" mechanism is not generally applicable, and one must develop an alternative approach. We show below that the nematic bond theory, which determines the exchange coupling bonds self-consistently using the $1/N$ expansion, provides a powerful framework that is universally applicable.

Our starting point is the path integral formulation of the partition function $Z=\int\mathcal{D}[\bf{S}_j]\delta(|\bf{S}_j|^2-1)e^{-\beta H}$, where $\beta=1/T$. With the help of the constraint field $\lambda_j$ (Lagrange multiplier), $Z$ may be written as
\begin{equation}
\label{eq:Z}
Z=\int \mathcal{D}[\bf{S}_j,\lambda_j]e^{-S},\,\,\, S=\beta H+i\sum_j\lambda_j(|\bf{S}_j|^2-1).
\end{equation}
Now that the action $S$ is quadratic in the spin variables, the functional integral over $\bf{S}_j$ can be performed exactly. The action for the constraint field is then 
\begin{equation}
\label{eq:S1}
S_\lambda=\frac{N}{2}{\rm tr}\log\left(\hat{J}+iT\hat{\lambda}\right)-i\sum_j\lambda_j,
\end{equation}
where $\hat{J}_{\bf{q}\bf{q}^\prime}=J_\bf{q}\delta_{\bf{q}\bf{q}^\prime}$ and $\hat{\lambda}_{\bf{q}\bf{q}^\prime}=\lambda_{\bf{q}-\bf{q}^\prime}/\sqrt{V}$ are matrices in momentum space, and $V$ is the system volume (we set the lattice spacing to unity). The susceptibility $\chi_\bf{q}$ is found by adding a source field action $S_\bf{h}=\bf{h}_\bf{q}\cdot \bf{S}_\bf{q}$ to Eq.~\eqref{eq:Z}, giving 
\begin{equation}
\label{eq:chi1}
\chi_\bf{q}=\frac{NT}{2}\left\langle\left(\hat{J}+iT\hat{\lambda}\right)_{\mathbf{qq}}^{-1}\right\rangle.
\end{equation} 
Here the average is taken with respect to the action $S_\lambda$, Eq.~\eqref{eq:S1}. We now consider the $1/N$ expansion (for a review see e.g. Refs.~\cite{Chubukov94,Sachdev-book}). At leading order in $1/N$, the action $S_\lambda$ can be evaluated at its saddle point. This leads to the equations $\lambda_{\rm saddle}=-i\beta\Delta$ and
\begin{equation}
\label{eq:con1}
\frac{N}{2}\int_\bf{q}\frac{T}{J_\bf{q}+\Delta}=\int_\bf{q}\chi_\bf{q}=1,
\end{equation}
where $\int_\bf{q}$ denotes a Brillouin zone integral. One may notice that the second equality in Eq.~\eqref{eq:con1} is nothing but the local constraint $|\bf{S}_j|^2=1$ written in $\bf{q}-$space. The parameter $\Delta$ is fixed by Eq.~\eqref{eq:con1}, and has a positive solution at all $T>0$. This implies the presence of a finite spin correlation length $\xi\propto1/\sqrt{\Delta}$, in accordance with the Mermin-Wagner theorem. Notably, the exchange bonds in Eq.~\eqref{eq:con1}, $J^{\rm eff}_\bf{q}=J_\bf{q}$, are trivially point group symmetric at all $T>0$.

%%%%%%%%%%%
%%%%%%%%%%%
\begin{figure}[t]
\includegraphics[width=.9\columnwidth]{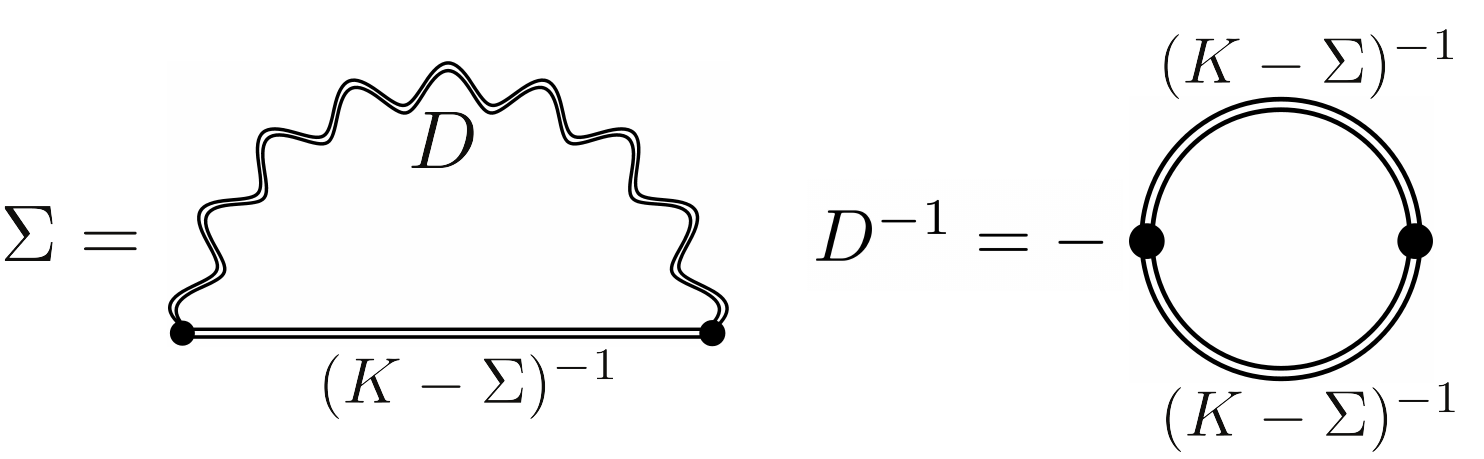}
\caption{Left: Leading order self-consistent self-energy diagram corresponding to Eq.~\eqref{eq:Sigma}. Right: The inverse constraint propagator $D^{-1}$ is itself a pure self-energy that contains, but is distinct from, the spin self-energy $\Sigma$ implicitly, see Eq.~\eqref{eq:D}.}
\label{fig:2}
\end{figure}
%%%%%%%%%%%
%%%%%%%%%%%

Fluctuations around the saddle point $\lambda_{\rm saddle}$ generate corrections to $\chi_\bf{q}$ via the simultaneous expansion of Eqs.~(\ref{eq:S1}-\ref{eq:chi1}) with respect to $\delta\lambda=\lambda-\lambda_{\rm saddle}$. To second order in $\delta\lambda$ Eq.~\eqref{eq:S1} becomes
\begin{equation}
\label{eq:S2}
S_\lambda^{(2)}=\sum_\bf{p}\frac{|\delta\lambda_\bf{p}|^2}{2D^{(0)}_\bf{p}},D^{(0)}_\bf{p}=\frac{2}{NT^2}\left(\int_\bf{k}\frac{1}{K_\bf{k}}\frac{1}{K_{\bf{k}+\bf{p}}}\right)^{-1},
\end{equation}
where $K_\bf{q}=J_\bf{q}+\Delta$. The second order term in the $\delta\lambda$ expansion of Eq.~\eqref{eq:chi1} can then be calculated using Eq.~\eqref{eq:S2}. This yields the spin self-energy correction $\Sigma^{(1)}_\bf{q}=-T^2\int_\bf{p}D^{(0)}_\bf{p}K^{-1}_{\bf{p}+\bf{q}}$ to the exchange bonds $J_\bf{q}^{\rm eff}=J_\bf{q}-\Sigma_\bf{q}^{(1)}$. As a result, the susceptibility becomes $\chi_\bf{q}=\frac{NT}{2}(K_\bf{q}-\Sigma_\bf{q}^{(1)})^{-1}$, where $\Delta$ must be chosen to satisfy the sum rule $\int_\bf{q}\chi_\bf{q}=1$ (note that $\chi^{-1}_\bf{q}$ is no longer a linear function of $\Delta$). Higher order terms in Eqs.~(\ref{eq:S1}-\ref{eq:chi1}) generate a power series in $1/N$ for the self-energy $\Sigma_\bf{q}$ \cite{Chubukov94,Sachdev-book}. For any finite order $m$ in $1/N$, $\Sigma_\bf{q}^{(m)}$ (and thus $\chi_\bf{q}^{(m)}$) will preserve the point group symmetry inherent in $J_\bf{q}$ at all $T$, such that $J_\bf{q}^{\rm eff}$ remains symmetric. In order for the exchange bonds to develop spontaneous symmetry breaking of the point group, one must thus perform an infinite order resummation. To this end, we determine the exchange bonds self-consistently at leading order in $1/N$. This yields the self-consistent bond equation
\begin{eqnarray}
\label{eq:Sigma}
K_\bf{q}^{\rm eff}&=& K_\bf{q}+\int_\bf{p}\frac{T^2D_\bf{p}}{K^{\rm eff}_{\bf{p}+\bf{q}}}.
\end{eqnarray}
This equation is depicted diagrammatically in Fig.~\ref{fig:2} for the spin self energy $\Sigma$, which is related to the exchange bonds via $K^{\rm eff}_\bf{q}=K_\bf{q}-\Sigma_\bf{q}$. Importantly, in Eq.~\eqref{eq:Sigma} the constraint propagator $D$ also contains $K_\bf{q}^{\rm eff}$ (cf. Fig.~\ref{fig:2}), and is given by
\begin{eqnarray}
\label{eq:D}
D_\bf{p}&=&\frac{2}{NT^2}\left(\int_\bf{k}\frac{1}{K^{\rm eff}_\bf{k}}\frac{1}{K^{\rm eff}_{\bf{k}+\bf{p}}}\right)^{-1}.
\end{eqnarray}
Due to the $1/T^2$ factor in $D_\bf{p}$, the explicit $T$ dependence in Eq.~\eqref{eq:Sigma} drops out, allowing one to parametrize solutions $K^{\rm eff}_\bf{q}(\Delta)$ using $\Delta$ only. The solution's temperature can then be determined once $K^{\rm eff}_\bf{q}(\Delta)$ is known using the sum rule $\int_\bf{q}\chi_\bf{q}=1$ or $T(\Delta)=(\frac{N}{2}\int_\bf{q}\frac{1}{K_\bf{q}^{\rm eff}(\Delta)})^{-1}$. Inverting $T(\Delta)\to\Delta(T)$ finally gives $\chi_\bf{q}(T)=NT/(2K^{\rm eff}_\bf{q}(T))$.

Some remarks regarding Eqs.~(\ref{eq:Sigma}-\ref{eq:D}) are in order. First, to obtain the spin self-energy $\Sigma$ within $D$, Eq.~\eqref{eq:D}, it is necessary to perform an infinite order resummation using all nonlinear terms $\delta\lambda^n$ generated by the logarithm in Eq.~\eqref{eq:S1}. In this way, one may replace the bare spin propagator $K$ with $K-\Sigma=K^{\rm eff}$ in Eq.~\eqref{eq:D}. Obtaining $K^{\rm eff}$ in the denominator of Eq.~\eqref{eq:Sigma} follows from the standard non-crossing approximation. Additional vertex and $\lambda$ self-energy corrections on top of Eqs.~(\ref{eq:Sigma}-\ref{eq:D}), are subleading in $1/N$ and can be neglected.

We now show that Eqs.~(\ref{eq:Sigma}-\ref{eq:D}) are sufficient to support the spontaneous formation of long-range nematic order. We write $\Sigma_\bf{q}=\Sigma_\bf{q}^{\rm e}-\sigma_\bf{q}$, where $\sigma_\bf{q}$ is the symmetry breaking component of $K^{\rm eff}_\bf{q}$ and $\Sigma^{\rm e}$ is a symmetry preserving renormalization of $K_\bf{q}$. Expanding Eq.~\eqref{eq:Sigma} to linear order in $\sigma_\bf{q}$ yields the equation for the critical point
\begin{equation}
\label{eq:Tc1}
\sigma_\bf{q}=\int_{\bf{q}^\prime}\frac{\sigma_{\bf{q}^\prime}}{\tilde{K}_{\bf{q}^\prime}^2}\left(-T^2D_{\bf{q}^\prime-\bf{q}}+\int_\bf{p}\frac{NT^4 D_\bf{p}^2}{\tilde{K}_{\bf{p}+\bf{q}}\tilde{K}_{\bf{p}+\bf{q}^\prime}}\right),
\end{equation}
where $\tilde{K}_\bf{q}=K_\bf{q}-\Sigma^{\rm e}_\bf{q}$ and now only $\Sigma^{\rm e}$ enters $D$. Equation~\eqref{eq:Tc1} can be understood as a matrix equation $\sum_{\bf{q}^\prime}(\delta_{\bf{q}\bf{q}^\prime}-\mathcal{M}_{\bf{q}\bf{q}^\prime})\sigma_{\bf{q}^\prime}=0$, whose matrix elements are labeled by $\bf{q}$. A nontrivial zero-mode, $\sigma_\bf{q}\neq0$, that breaks point group symmetry signals the onset of nematic order.

We now consider the simplest case of an Ising-nematic critical point where there is a two-fold degenerate groundstate with wavevectors $\bf{Q}_\alpha$, $\alpha=(1,2)$ that are equivalent under rotation. For small $\Delta$ (and thus $T$), the factor $\tilde{K}^{-2}_{\bf{q}^\prime}$ in Eq.~\eqref{eq:Tc1} is strongly peaked at $\bf{q}^\prime=\pm\bf{Q}_{1,2}$, allowing one to evaluate the remaining terms at such points. Setting $\bf{q}=\bf{Q}_1$ in Eq.~\eqref{eq:Tc1} gives
\begin{equation}
\label{eq:Tc2}
1=g_2\int_\bf{q}\frac{1}{K_\bf{q}^2}.
\end{equation}
Equation~\eqref{eq:Tc2} provides the condition for a zero eigenvalue of the nematic mode $(\sigma_{\bf{Q}_1},\sigma_{\bf{Q}_2})=\sigma(1,-1)$, where $\sigma$ is the order parameter amplitude.
In Eqs.~(\ref{eq:Tc2}) and (\ref{eq:g}) below, we neglect the symmetric component $\Sigma^{\rm e}\propto 1/N$ compared to $K_\bf{q}$. In cases where $J_\bf{q}$ develops line nodes, however, the term $\Sigma^{\rm e}$ is important for lifting the accidental continuous degeneracy. In Eq.~\eqref{eq:Tc2} $g_2$ is 
\begin{eqnarray}
\nonumber g_2&=&\frac{NT^4}{16}\int_\bf{p}D_\bf{p}^2\left(\sum\limits_{s=\pm1}(K_{\bf{p}+s\bf{Q}_1}^{-1}-K_{\bf{p}+s\bf{Q}_2}^{-1})\right)^2
\\
\label{eq:g}&+&\frac{T^2}{4}(D_{\bf{Q}_1+\bf{Q}_2}+D_{\bf{Q}_1-\bf{Q}_2}-D_{2\bf{Q}_1}-D_0).
\end{eqnarray}
The form of $g_2$ in Eq.~\eqref{eq:g} implies that near a rotation symmetric Lifshitz point $\bf{Q}_0$ (e.g. $\bf{Q}_0=(0,0)\,\, {\rm or}\, (\pi,\pi)$ on the square lattice), $g_2\propto |\bf{Q}-\bf{Q}_0|^4$ as $\bf{Q}\to\bf{Q}_0$. A scaling analysis of Eq.~\eqref{eq:Tc2} for $T$ and $|\bf{Q}|$ gives $1=g_2\int_\bf{q}K_\bf{q}^{-2}\propto |\bf{Q}-\bf{Q}_0|^4/\Delta$, where $\Delta\propto T^2$ at $\bf{Q}=\bf{Q}_0$ from Eq.~\eqref{eq:con1}. We thus obtain \begin{equation}
\label{eq:Lifshitz}
T_c\propto |\bf{Q}-\bf{Q}_0|^2,\,{\rm as}\,\,\bf{Q}\to\bf{Q}_0.
\end{equation}
While the prefactor to Eq.~\eqref{eq:Lifshitz} is model dependent, the wavevector scaling of $T_c$ is universal.  Remarkably, $T_c$ is parametrically \emph{larger} than the intuitive estimate \cite{Villain77} given by the exchange energy barrier $W\propto |\bf{Q}-\bf{Q}_0|^{4}$ separating the two degenerate groundstates.

Beyond the Ising-nematic case, Eq.~\eqref{eq:Tc1} also applies to an arbitrary $n-$state Potts-nematic critical point, associated with $n-$fold groundstate degeneracy. The case $n=3$ occurs, e.g., in the frustrated Heisenberg ferromagnet on a triangular lattice with next-nearest-neighbor antiferromagnetic exchange coupling $J_2>|J_1|/3$.  In 2D the 3-state Potts transition is continuous \cite{Wu82} (of course, this does not guarantee a continuous transition in the frustrated Heisenberg system), and one may look for the zero-mode $(\sigma_{\bf{Q}_1},\sigma_{\bf{Q}_2},\sigma_{\bf{Q}_3})=\sigma(1,-\frac{1}{2},-\frac{1}{2})$. The instability condition is given by Eq.~\eqref{eq:Tc2} with $g_2$ replaced by $g_3$ for $n=3$ (the precise form of $g_3$ is not important here and will be discussed elsewhere). The scaling analysis preceeding Eq.~\eqref{eq:Lifshitz}, and thus Eq.~\eqref{eq:Lifshitz} itself, holds for an arbitrary $n$.

It is important to emphasize an additional possibility allowed by Eq.~\eqref{eq:Tc1}: that of a rotation \emph{symmetric} state that breaks lattice mirror symmetries. This may occur in a regime of strong frustration where $J_\bf{q}$ (or $J^{\rm eff}_\bf{q}$) develops minima at wavevectors that are not invariant under lattice mirror reflections. Since the groundstates still break rotation symmetry, however, such a phase is expected to transition into a nematic state at sufficiently low $T$.
An example is the recently discovered  magnetic vortex-antivortex phase \cite{Seabra16} of the ferromagnetic $J_1$-$J_2$-$J_3$ model (which preserves rotation symmetry but not lattice mirror symmetries). As expected, this phase exists in an intermediate temperature range in the vicinity of strong frustration $J_3\sim J_2/2$, but we will not address it further here.

%%%%%%%%%%%
%%%%%%%%%%%
\begin{figure}[t]
\includegraphics[width=\columnwidth]{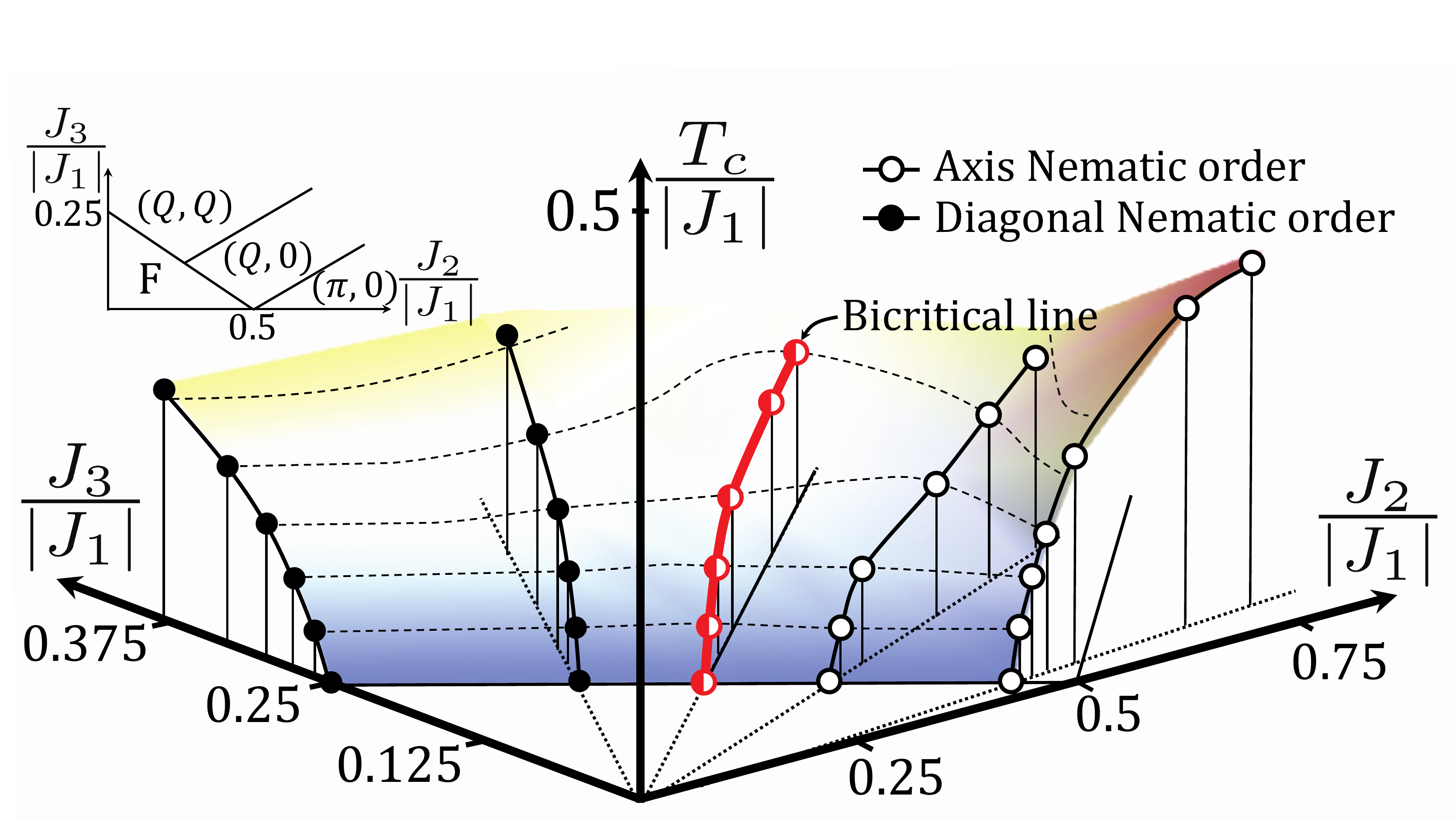}
\caption{(Color online) Schematic phase diagram of the square-lattice ferromagnetic $J_1$-$J_2$-$J_3$ model, Eq.~\eqref{eq:J123}. Inset: groundstate phase diagram with the helimagnetic wavevector $\textbf{Q}=(Q_x,Q_y)$ indicated. As one approaches the ferromagnetic (F) phase the spin-nematic transition temperature vanishes linearly. Open(solid) circles denote the onset of spin-nematic order $\sigma_a(\sigma_d)$ directed along the $x/y$ axes (diagonal/anti-diagonal). The intersection of critical surfaces, $T_c^{(a)}=T_c^{(d)}$, leads to a line of bicritical points.}
\label{fig:3}
\vspace{-3mm}
\end{figure}
%%%%%%%%%%%
%%%%%%%%%%%

We now apply our results for the Ising-nematic transition to the square lattice ferromagnetic $J_1$-$J_2$-$J_3$ model. The Hamiltonian is defined in Eq.~\eqref{eq:H} with the exchange coupling
\begin{equation}
\label{eq:J123}
J_\bf{q}=-J_1\left(c_x+c_y\right)+2 J_2 c_x c_y+2J_3 \left(c_x^2 +c_y^2\right)-J_0,
\end{equation}
where $c_\eta=\cos q_\eta$, $\eta=x,y$, $J_{1,2,3}>0$ and $J_0$ is a constant shift chosen to have a zero energy groundstate (${\rm min}\,J_\bf{q}=0$). For $J_\bf{q}$ in Eq.~\eqref{eq:J123} the system posseses four distinct groundstate phases (inset of Fig.~\ref{fig:3}): a ferromagnetic (F) phase for $J_3<1/4-J_2/2$ (we set $J_1=1$), generally incommensurate helicoidal phases for $J_3>|1/4-J_2/2|$ with $\bf{Q}_\alpha$ parallel to the lattice axes $(J_3<J_2/2)$ or lattice diagonals $(J_3>J_2/2)$. A commensurate phase occurs for $J_3<J_2/2-1/4$.

The nematic transition temperature is calculated numerically from Eqs.~(\ref{eq:Tc2}-\ref{eq:g}). A qualitative sketch of the resulting phase diagram is given in Fig.~\ref{fig:3}. Two essential features of the phase diagram can be observed: (i) $T_c$  vanishes \emph{linearly} with exchange parameters as the ferromagnetic phase is approached and (ii) there exists an intersection of two nematic critical surfaces along a bicritical line that runs along $J_3\approx J_2/2$. As we show below (i) follows directly from Eq.~\eqref{eq:Lifshitz}. (ii) Occurs roughly when the groundstate wavevector changes from axis to diagonal alignment through a 1st order Lifshitz transition of $J_\bf{q}$, which develops a circular line node.
\begin{comment}
Phenomenologically, the action near the bicritical line can be written as
\begin{equation}
\label{eq:sigma}
S=\frac{1}{2}(r_a\sigma_a^2+r_d\sigma_d^2)+\frac{1}{4}b_a\sigma_a^4+b_d\sigma_d^4+\frac{1}{2}b_{ad} \sigma_a^2 \sigma_d^2,
\end{equation}
\end{comment}

Preliminary Monte Carlo data \footnote{M. Schecter, O. F. Sylju{\aa}sen and J. Paaske, (unpublished).} in the vicinity of the bicritical point strongly suggest 2nd order thermal phase transitions into pure phases below $T_c$, with either $\sigma_ a \neq 0, \sigma_d=0$ or $\sigma_a=0, \sigma_d \neq 0$, i.e. no phase coexistence. Here $\sigma_{a\{d\}} =\int_\bf{q}\sigma_\bf{q}f^{a\{d\}}_\bf{q}$ with
$f^{a\{d\}}_\bf{q}=(\cos q_x-\cos q_y)\{(\cos \frac{q_x+q_y}{2}-\cos \frac{q_x-q_y}{2})\}$.
For fixed $T$ there is thus a 1st order transition from axis to diagonal nematic order (or vice versa) as one varies the exchange parameters. A nonlinear analysis of Eq.~\eqref{eq:Sigma} is required to determine whether this behavior also occurs within the self-consistent bond theory.

\begin{figure}
  \includegraphics[width=\columnwidth]{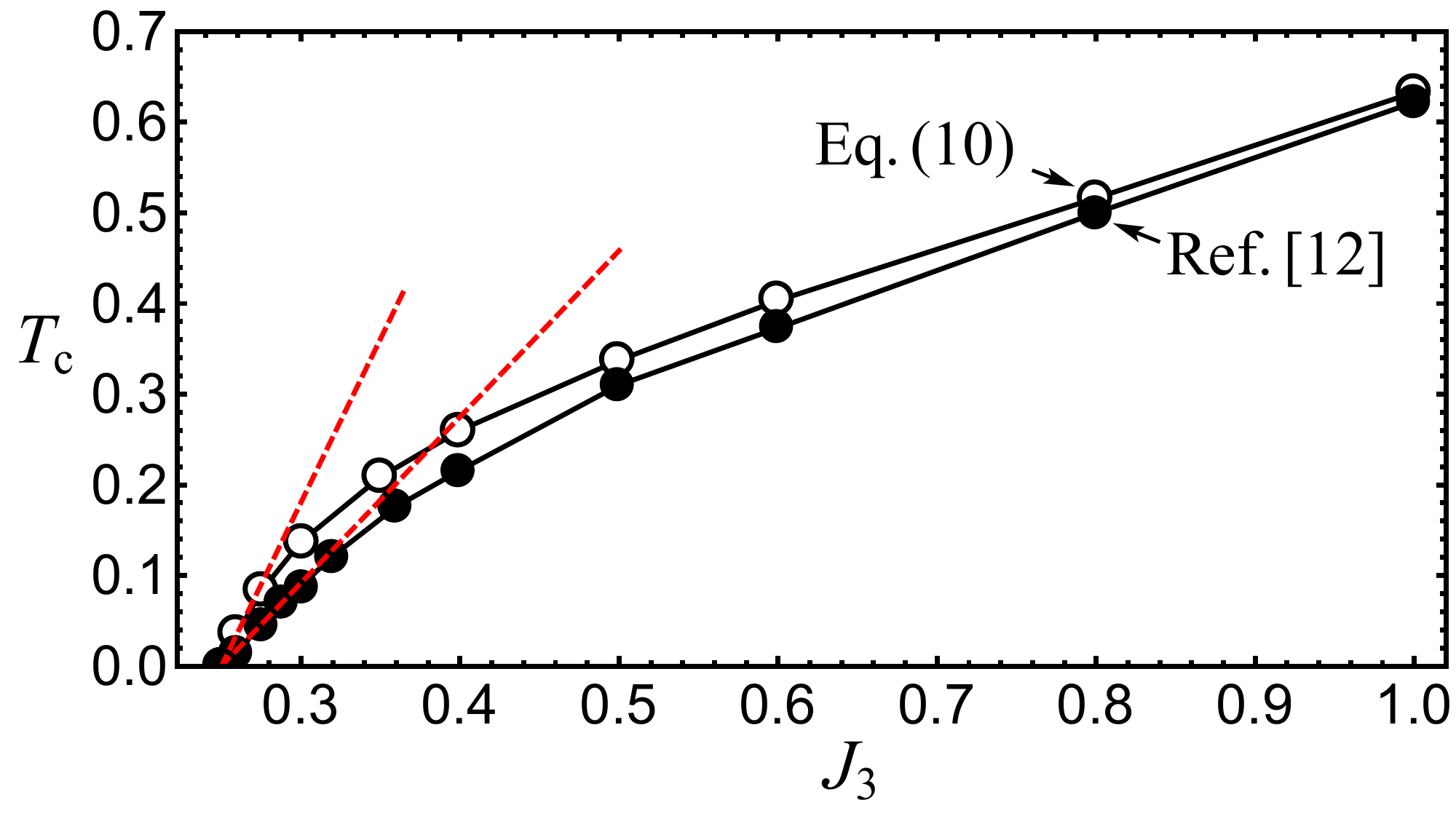}
  \caption{(Color online) $T_c$ versus $J_3$ for $J_2=0$, comparing Eq.~\eqref{eq:Tc2} and Monte Carlo simulations from Ref.~\cite{Capriotti04}. The linear vanishing of $T_c$ as $J_3\searrow 1/4$ is indicated by the red dashed lines, and is consistent with the general result Eq.~\eqref{eq:Lifshitz}.}
\label{fig:5}
\vspace{-4mm}
\end{figure}

Finally, the linear vanishing of $T_c$ with exchange parameters follows trivially from Eq.~\eqref{eq:Lifshitz} once the relation to the wavevector $\bf{Q}$ is established. We find
\begin{equation}
\label{eq:Lifshitz2}
T_{c}\propto|\mathbf{Q}|^{2}\approx\begin{cases}
8(2J_{3}+J_{2}-1/2), & J_{3}>1/8,\\
\frac{2J_{3}+J_{2}-1/2}{J_{3}}, & J_{3}<1/8,
\end{cases}
\end{equation}
where $J_{3}\approx1/4-J_{2}/2$. The result Eq.~\eqref{eq:Lifshitz2} was first observed in Monte Carlo simulations for $J_2=0$ \cite{Capriotti04}, where $T_c\propto J_3-J_1/4$ near the Lifshitz point $J_3=J_1/4$, cf. Eq.~\eqref{eq:Lifshitz2}. This behavior can be seen in Fig.~\ref{fig:5}, which compares the results of Eq.~\eqref{eq:Tc2} and Ref.~\cite{Capriotti04}, showing good agreement. Figure~\ref{fig:3} demonstrates that the scaling law Eq.~\eqref{eq:Lifshitz2} holds along the entire line of Lifshitz points $J_3=1/4-J_2/2$, $J_2<1/2$. This scaling does not apply at the point $J_3=0,J_2=1/2$ because the peak wavevector near $T=T_c$ does not evolve continuously from $\bf{Q}=(0,\pi)$ to $\bf{Q}=0$ as $J_2\searrow1/2$ at $J_3=0$; peak height is transferred between $\bf{Q=0}$ and $(0,\pi)$ in a first order manner as $T$ is lowered \cite{Weber03}, similar to having a thermally generated real space $J_3^{\rm eff}<0$ term in the susceptibility. This leads to the vanishing of $T_c$ as $J_2\searrow1/2$ with an infinite slope \cite{Weber03}, fully consistent with Eq.~\eqref{eq:Lifshitz2}, since the slope diverges as $1/J_3$ as $J_3\to 0$.  These features ($T_c\to0$ and $J_3^{\rm eff}<0$) are at odds with linear spin-wave theory \cite{Henley89,Chandra90} which predicts a \emph{peak} in $T_c$ at $J_2=1/2$ and $J_3^{\rm eff}>0$ \cite{Chandra90}. This indicates the prominence of nonlinear spin-wave effects, which are included nonperturbatively in Eq.~\eqref{eq:Sigma}.

%%%%%%%%%%%
%%%%%%%%%%%
%%%%%%%%%%%
%%%%%%%%%%%

In conclusion, we have developed a theory of nematic spin order in 2D Heisenberg helimagnets based on a self-consistent determination of the exchange bond couplings. This nematic bond theory naturally preserves the spin-rotation symmetry, and describes nematic order in systems with helicoidal correlations. The critical temperature was found to depend sensitively on the peak helimagnetic wavevector, and vanishes continuously when approaching rotation symmetric Lifshitz points. Generalizations to incorporate a finite magnetic field or quantum spins are interesting directions for the future.

\acknowledgements{We thank B. M. Andersen for discussions and comments on the manuscript. The Center for Quantum Devices is funded by the Danish National Research Foundation. Support from the Villum Foundation (MS) and grant NFR213606 (OFS) is acknowledged.}

\bibliography{bib-schecter17}

\end{document}